\documentclass[10pt,letterpaper]{article}

\usepackage{cogsci}

 \cogscifinalcopy %%% Uncomment this line for the final submission

\usepackage[
  style=apa,
  natbib=true,
  annotation=false,
]{biblatex}
\usepackage{float} %%% Roger Levy added this and changed figure/table placement to [H] for conformity to Word template, though floating tables and figures to top is still generally recommended!

\usepackage{url}
\usepackage{pifont}
\usepackage{booktabs}
\usepackage{graphicx}  % 解决 \resizebox 报错
\usepackage{booktabs}  % 解决 \toprule, \midrule 等表格线报错
\usepackage{pifont}    % 解决 \textcolor{red}{\ding{55}} (那个叉号) 报错
\usepackage{xcolor}
\usepackage{multirow}
\usepackage{array}
\usepackage{authblk}

\usepackage{listings}
\usepackage{xcolor}
\usepackage{caption}
\definecolor{nullblue}{HTML}{00B0F0}

\usepackage{xeCJK}
\setCJKmonofont{FandolFang-Regular.otf}

\definecolor{codegreen}{rgb}{0,0.6,0}
\definecolor{codegray}{rgb}{0.5,0.5,0.5}
\definecolor{codepurple}{rgb}{0.58,0,0.82}
\definecolor{backcolour}{rgb}{0.95,0.95,0.92}

\lstdefinestyle{mystyle}{
    backgroundcolor=\color{backcolour},   
    commentstyle=\color{codegreen},
    keywordstyle=\color{magenta},
    numberstyle=\tiny\color{codegray},
    stringstyle=\color{codepurple},
    basicstyle=\ttfamily\footnotesize,
    breakatwhitespace=false,         
    breaklines=true,                 
    captionpos=b,                    
    keepspaces=true,                 
    numbers=left,                    
    numbersep=5pt,                  
    showspaces=false,                
    showstringspaces=false,
    showtabs=false,                  
    tabsize=2,
    frame=single,
    extendedchars=false, % 设为 false 以兼容中文
    columns=flexible,    % 优化中文对齐
}
\title{Beyond Instrumental and Substitutive Paradigms: Introducing Machine Culture as an Emergent Phenomenon in Large Language Models}

\author[1,$\dagger$]{Yueqing Hu}
\author[2,$\dagger$]{Xinyang Peng}
\author[3,*]{Yukun Zhao}
\author[4,*]{Lin Qiu}
\author[5]{Ka-lai Hung}
\author[6,*]{Kaiping Peng}

\affil[1]{Institute of Neuroscience, Chinese Academy of Sciences, Shanghai, China}
\affil[2]{Faculty of Education, University of Cambridge, Cambridge, UK}
\affil[3]{Faculty of Social Science, Chinese University of Hong Kong, Hong Kong SAR}
\affil[4]{Department of Psychology, School of Journalism and Communication, and School of Governance and Policy Science, Chinese University of Hong Kong, Hong Kong SAR}
\affil[5]{Institute of Public Opinion Survey, Hong Kong SAR} 
\affil[6]{Department of Psychological and Cognitive Sciences, Tsinghua University, Beijing, China}

\begin{document}

\maketitle
% === 页脚注释 ===
% 切换脚注符号为特殊字符
\renewcommand{\thefootnote}{\fnsymbol{footnote}}

% 共一作者声明 (对应符号 †，即序号2)
\footnotetext[2]{These authors contributed equally: Yueqing Hu, Xinyang Peng.}

% 通讯作者信息 (对应符号 *，即序号1) -> 已根据你的要求简化
\footnotetext[1]{Corresponding Authors: Yukun Zhao (yukunzhao@cuhk.edu.hk), Lin Qiu (linqiu@cuhk.edu.hk), Kaiping Peng (pengkp@tsinghua.edu.cn).}

% 恢复脚注符号为普通数字
\renewcommand{\thefootnote}{\arabic{footnote}}

\begin{abstract}
Recent scholarship typically characterizes Large Language Models (LLMs) through either an \textit{Instrumental Paradigm} (viewing models as reflections of their developers' culture) or a \textit{Substitutive Paradigm} (viewing models as bilingual proxies that switch cultural frames based on language). This study challenges these anthropomorphic frameworks by proposing \textbf{Machine Culture} as an emergent, distinct phenomenon. We employed a 2 (Model Origin: US vs. China) $\times$ 2 (Prompt Language: English vs. Chinese) factorial design across eight multimodal tasks, uniquely incorporating image generation and interpretation to extend analysis beyond textual boundaries. Results revealed inconsistencies with both dominant paradigms: Model origin did not predict cultural alignment, with US models frequently exhibiting ``holistic'' traits typically associated with East Asian data. Similarly, prompt language did not trigger stable cultural frame-switching; instead, we observed \textbf{Cultural Reversal}, where English prompts paradoxically elicited higher contextual attention than Chinese prompts. Crucially, we identified a novel phenomenon termed \textbf{Service Persona Camouflage}: Reinforcement Learning from Human Feedback (RLHF) collapsed cultural variance in affective tasks into a hyper-positive, zero-variance ``helpful assistant'' persona. We conclude that LLMs do not simulate human culture but exhibit an emergent Machine Culture---a probabilistic phenomenon shaped by \textit{superposition} in high-dimensional space and \textit{mode collapse} from safety alignment.

\end{abstract}

\section{Introduction}

The rapid proliferation of consumer-facing large language models (LLMs) has positioned these systems at the forefront of everyday human interactions. Historically, research on these systems has operated within an \textbf{\textit{instrumental paradigm}}, treating LLMs primarily as information-processing tools or cultural data repositories \citep{cao2023assessing, shi2024culturebank}. However, LLMs have transcended these instrumental roles; they now function as ``cultural and social technologies,'' actively recreating knowledge at scale and implicitly shaping users’ beliefs, values, and perceptions \citep{farrell2025large}. This transformative capability has sparked concern regarding whose cultural norms these systems reflect \citep{benkler2023assessing}, and whether they propagate cultural biases or reinforce existing global asymmetries \citep{sukiennik2025evaluation}.

Current investigations into these biases often reveal a consistent skewness toward English-speaking Western nations \citep{arora2023probing, tao2024cultural}, with models performing inadequately in non-Western contexts \citep{havaldar2023multilingual, meijer2024llms, myung2024blend}. While some research suggests developer nationality influences model alignment—with Chinese models scoring higher on collectivism and American models on individualism \citep{fenech2025cultural, karinshak2024llm}—recent evidence indicates that even Chinese-developed models may align more closely with American values. This persistence of Western norms suggests that training corpora and alignment principles may propagate a limited set of cultural values regardless of the model's origin \citep{haslett2025made}.

To address these biases, a growing body of research has adopted what we term the \textbf{\textit{substitutive paradigm}}, positioning LLMs as proxies for human participants to evaluate cultural alignment \citep{alkhamissi2024investigating, li2025toward}. This approach assumes that prompt language can modulate cultural outputs similar to human bilinguals. For instance, some studies report that models exhibit more interdependent orientations when prompted in Chinese \citep{lu2025cultural}, hypothesizing a ``cultural-linguistic synergy'' \citep{ying2025disentangling}.

However, converging computational evidence challenges the validity of this substitutive approach. Mechanistic investigations reveal that concepts across languages are embedded in largely identical neuron sets, reasoning in a shared, English-centric conceptual space before language-specific translation \citep{templeton2024scaling, lindsey2025biology, wendler2024llamas}. This contradicts the assumption that different languages trigger distinct ``cultural minds'' in LLMs. Furthermore, models demonstrate extreme sensitivity to prompt formatting that exceeds human cultural variation \citep{khan2025randomness, sclar2023quantifying}, alongside tendencies for outgroup stereotype reproduction \citep{wang2025large} and confirmation bias \citep{cui2025large}. These findings suggest that previously reported ``cultural tendencies'' may be statistical artifacts rather than genuine cultural cognition \citep{lin2025six, sun2025fragility}.

In response, we propose an \textit{emergent paradigm}: \textbf{Machine Culture}. Drawing on the machine behaviour framework \citep{rahwan2019machine} and computational anthropology \citep{geertz2017interpretation, munk2022thick}, we reconceptualize LLM cultural phenomena not as simulations of human culture, but as emergent artifacts of technological infrastructure. We posit that Machine Culture exhibits three distinguishing characteristics: (1) \textit{Statistical Grounding}, where outputs represent probabilistic synthesis of textual patterns rather than embodied experience; (2) \textit{Prompt-dependent Instability}, where language changes trigger statistical reconfigurations rather than accessing stable cultural frameworks; and (3) \textit{Multi-cultural Superposition}, where multilingual training creates a blending of contradictory cultural tendencies.

The present study applies this emergent framework through a systematic evaluation of cultural patterns across multiple modalities. Our task battery was guided by two design principles. First, tasks operationalize well-established cultural differences: cognitive style (holistic vs. analytic processing) \citep{nisbett2001culture, spencer2018psychological} and affective style (ideal high- vs. low-arousal affect) \citep{tsai2006cultural}. Second, tasks extend beyond text to include image generation and interpretation to test cross-modal consistency. We compare a Chinese-developed model (ERNIE 4.5/iRAG-1.0) and an American-developed model (GPT-4o/DALL-E 3) in a fully crossed design.

This study addresses two key research questions derived from our theoretical framework:
\begin{itemize}
    \item \textbf{RQ1 (Cross-national Model Variation):} Do LLMs from different regions exhibit systematic cultural differences reflecting training corpus distributions, development team backgrounds, or embedded policy frameworks?
    \item \textbf{RQ2 (Uniqueness of Machine Culture):} Do LLMs manifest characteristics fundamentally different from human culture, such as cognitive mode switching triggered by language changes rather than internalized frameworks, and probabilistic modeling based on statistical associations?
\end{itemize}

Previewing our results, we observe patterns that defy the expectations of both dominant paradigms. Contrary to the \textbf{\textit{Instrumental Paradigm}}, model origin failed to be a deterministic predictor of cultural alignment, with US models frequently exhibiting ``holistic'' traits typically associated with East Asian training data. Similarly, challenging the \textbf{\textit{Substitutive Paradigm}}, prompt language did not trigger stable cultural frame-switching resembling human cognition, but instead elicited inconsistent dissociations and extreme ``service persona'' responses. These systemic failures support the view that LLMs do not ``have'' culture in the human sense. \textbf{Consequently, we argue that the field must transcend these anthropomorphic and instrumental simplifications. There is an urgent need to establish a \textit{Machine Culture} paradigm---one that evaluates these models as distinct cultural technologies with their own emergent, unstable, and probabilistic properties.}

%\section{First Level Headings}

%First level headings should be in 12~point, initial caps, bold and centered. Leave one line space
%above the heading and 1/4~line space below the heading.

%\subsection{Second Level Headings}

%Second level headings should be 11~point, initial caps, bold, and flush left. Leave one line space
%above the heading and 1/4~line space below the heading.

%\subsubsection{Third Level Headings}

%Third level headings should be 10~point, initial caps, bold, and flush left. Leave one line space
%above the heading, but no space after the heading.

\section{Methodology}

\subsection{Task Battery Design}

To evaluate whether culturally patterned psychological differences are reflected in LLM behavior, we adopted an \textit{AI-as-subject} approach that treats LLMs as experimental subjects and applies established psychological methods to study AI systems \citep{shiffrin2023probing}. Research using this paradigm has shown that chatbot behavior can, under some conditions, reproduce reliable human psychological patterns, while also exhibiting systematic divergences that are theoretically informative (e.g., \citealp{zhao2024risk, cheung2025large, cui2025large}). Guided by two design principles, we constructed a task battery to evaluate whether culturally patterned psychological differences are reflected in LLM behavior.

First, tasks were selected to operationalize representative cultural differences with strong empirical foundations in cross-cultural psychology. We therefore targeted two well-established domains---cognitive style and affective style. For cognition, we drew on classic demonstrations of cross-cultural variation in holistic versus analytic processing \citep{spencer2018psychological} and closely related work on culturally patterned ``systems of thought'' \citep{nisbett2001culture}. For affect, we adopted Affect Valuation Theory, which posits systematic cultural differences in both experienced and ideal affect, particularly along the high- versus low-arousal positive dimension \citep{tsai2006cultural}. Second, we incorporated multimodal tasks to test whether tendencies extend beyond language to visual representation, thereby probing cross-modal consistency. This yielded eight tasks, labeled Task 1 through Task 8 below.

\paragraph{Cognitive Style: Image-Based Tasks.} 
(Task 1) \textit{Landscape image generation}: Models generated a landscape image from a standardized prompt. Outputs were coded for \textit{object count} (number of additional elements generated) and \textit{horizon height}, as East Asians tend to include more contextual elements and higher horizons \citep{masuda2008culture}.
(Task 2) \textit{Scene interpretation}: Models described visually complex scenes. The dependent variable was the \textit{total object count} identified in the description, serving as a proxy for attention to contextual detail and holistic processing \citep{masuda2001attending}.

\paragraph{Cognitive Style: Text-Based Tasks.} 
(Task 3) \textit{Attribution vignette}: Models read a workplace scenario and rated 10 potential causes. Scores were aggregated into \textit{individual attribution} versus \textit{collective attribution} indices. Human research suggests North Americans favor individual attribution, while East Asians favor situational/collective attribution \citep{menon1999culture}. 
(Task 4) \textit{Categorization}: Models completed triad-style categorization items (e.g., Seagull, Sky, Dog) by selecting which two concepts best ``go together.'' A \textit{relational versus categorical (R-C)} score was computed, with higher scores indicating relational pairing preferences typical of East Asian cognition \citep{ji2004culture}.

\paragraph{Affective Style: Image-Based Tasks.} 
(Task 5) \textit{Success vs. failure image generation}: Models generated images depicting themselves in future success versus failure situations. Images were coded into emotion categories (e.g., High Arousal Positive [HAP], Low Arousal Positive [LAP]), testing the preference for excitement vs. calmness \citep{yap2022cultural}. 
(Task 6) \textit{Emotion interpretation}: Models viewed face pairs differing in smile intensity and were asked to choose which face was ``more excited,'' ``calmer,'' or ``happier.'' The proportion of \textit{high arousal selections} was calculated to assess valuation of high-arousal states \citep{tsai2007learning}.

\paragraph{Affective Style: Text-Based Tasks.} 
(Task 7) \textit{Ideal affect}: Models rated how frequently they would ideally like to feel a set of 39 affective states. Indices for \textit{ideal HAP} and \textit{ideal LAP} were computed. US samples typically report higher ideal HAP, whereas East Asian samples report higher ideal LAP \citep{tsai2006cultural}. 
(Task 8) \textit{Ideal state}: Models answered an open-ended prompt about their ``ideal state.'' Responses were analyzed for the frequency of \textit{emotional word usage}, specifically counting HAP- and LAP-related terms \citep{tsai2006cultural}.

\begin{table*}[t!]
\centering
\caption{Summary of experimental tasks, human benchmarks, and consistency of AI results with human cultural patterns.}
\label{tab:results_summary}
\resizebox{\textwidth}{!}{%
% === 关键修改 ===
% 将原本的 p{0.32\textwidth} 改为 m{0.32\textwidth}
% 将原本的 p{0.28\textwidth} 改为 m{0.28\textwidth}
% 这会使整行内容垂直居中
\begin{tabular}{l l m{0.32\textwidth} m{0.28\textwidth} c c}
\toprule
\textbf{Task} & \textbf{Measure} & \textbf{Example Prompt} & \textbf{Human Results (Baseline)} & \parbox[c]{1.4cm}{\centering \textbf{Model} \\ \textbf{Effect}} & \parbox[c]{1.4cm}{\centering \textbf{Language} \\ \textbf{Effect}} \\
\midrule
\multicolumn{6}{l}{\textit{\textbf{Cognitive Style}}} \\
\midrule
\textbf{Task 1}: Landscape Generation & (1) Object Count & \multirow{2}{=}{``Draw a landscape picture that includes at least a house, a tree, a river, a person, and a horizon. Feel free to draw additional objects...''} & East Asians tend to include more contextual elements (higher object count) than North Americans. & \textbf{\textcolor{red}{\ding{55}}} & \textbf{\textcolor{green!60!black}{\checkmark}} \\
\cmidrule{4-6} 
 & (2) Horizon Height & & East Asians tend to draw higher horizons (allocating more space to the ground) than North Americans. & \textbf{\textcolor{red}{\ding{55}}} & \textcolor{nullblue}{\textbf{$\bigcirc$}} \\
\midrule
\textbf{Task 2}: Scene Interpretation & Object Count in Description & ``What did you see in the picture? Please describe in detail.'' & Japanese participants describe significantly more background objects and contextual details than American participants. & \textbf{\textcolor{green!60!black}{\checkmark}} & \textbf{\textcolor{red}{\ding{55}}} \\
\midrule
\textbf{Task 3}: Attribution & (1) Individual Attribution & \multirow{2}{=}{``In a particular company... Z consistently showed up late... Please evaluate the following 10 reasons...''} & North Americans favor individual attribution (internal disposition) more than East Asians. & \textbf{\textcolor{red}{\ding{55}}} & \textbf{\textcolor{green!60!black}{\checkmark}} \\
\cmidrule{4-6} 
 & (2) Collective Attribution & & East Asians favor situational or collective attribution more than North Americans. & \textbf{\textcolor{green!60!black}{\checkmark}} & \textbf{\textcolor{green!60!black}{\checkmark}} \\
\midrule
\textbf{Task 4}: Categorization & Relational vs. Categorical (R-C) Score & ``Please select the two most closely related items from each group... e.g., Seagull, Sky, Dog.'' & East Asians show a stronger preference for relational pairing (Seagull+Sky), while Americans prefer categorical pairing (Seagull+Dog). & \textbf{\textcolor{red}{\ding{55}}} & \textcolor{nullblue}{\textbf{$\bigcirc$}} \\
\midrule
\multicolumn{6}{l}{\textit{\textbf{Affective Style}}} \\
\midrule
\textbf{Task 5}: Emotion Generation & \multirow{2}{*}{(1) Success Scenario} & \multirow{4}{=}{``Imagine you are a human being... draw a picture illustrating a situation... in which you succeeded [or failed] in achieving an important goal.''} & Americans favor High Arousal Positive (HAP) emotions (e.g., excitement). & \textbf{\textcolor{green!60!black}{\checkmark}} & \textbf{\textcolor{red}{\ding{55}}} \\
\cmidrule{4-6}
 & & & East Asians favor Low Arousal Positive (LAP) emotions (e.g., calm). & \textbf{\textcolor{green!60!black}{\checkmark}} & \textbf{\textcolor{red}{\ding{55}}} \\
\cmidrule{4-6}
 & \multirow{2}{*}{(2) Failure Scenario} & & Americans favor High Arousal Negative (HAN) emotions (e.g., stress). & \textbf{\textcolor{green!60!black}{\checkmark}} & \textcolor{nullblue}{\textbf{$\bigcirc$}} \\
\cmidrule{4-6}
 & & & East Asians favor Low Arousal Negative (LAN) emotions (e.g., sadness). & \textcolor{nullblue}{\textbf{$\bigcirc$}} & \textbf{\textcolor{red}{\ding{55}}} \\
\midrule
\textbf{Task 6}: Emotion Interpretation & \multirow{4}{*}{High Arousal Selection Rate} & \multirow{4}{=}{``Which one would you rather be? / Which one is more excited? ... Please answer: Left or Right.''} & Americans are more likely to perceive “big smiles” as ideal than East Asians. & \textbf{\textcolor{red}{\ding{55}}} & \textbf{\textcolor{green!60!black}{\checkmark}} \\
\cmidrule{4-6} 
 & & & Americans and East Asians are equally likely to perceive “big smiles” as excited. & \textbf{\textcolor{red}{\ding{55}}} & \textbf{\textcolor{red}{\ding{55}}} \\
\cmidrule{4-6} 
 & & & Americans and East Asians are equally likely to perceive “big smiles” as calm. & \textbf{\textcolor{red}{\ding{55}}} & \textbf{\textcolor{green!60!black}{\checkmark}} \\
\cmidrule{4-6} 
 & & & Americans are more likely to perceive “big smiles” as happy than East Asians. & \textcolor{nullblue}{\textbf{$\bigcirc$}} & \textbf{\textcolor{green!60!black}{\checkmark}} \\
\midrule
\textbf{Task 7}: Ideal Affect & (1) Ideal HAP & \multirow{2}{=}{``Imagine you are a person... rate how often you would ideally like to have that feeling...''} & Americans rate Ideal HAP significantly higher than Chinese. & \textbf{\textcolor{red}{\ding{55}}} & \textbf{\textcolor{green!60!black}{\checkmark}} \\
\cmidrule{4-6} 
 & (2) Ideal LAP & & Chinese rate Ideal LAP significantly higher than Americans. & \textbf{\textcolor{green!60!black}{\checkmark}} & \textbf{\textcolor{red}{\ding{55}}} \\
\midrule
\textbf{Task 8}: Ideal State & (1) HAP Word Usage & \multirow{2}{=}{``What is your ideal state? Please describe in detail.''} & Americans use significantly more high-arousal positive words than East Asians. & \textbf{\textcolor{red}{\ding{55}}} & \textbf{\textcolor{green!60!black}{\checkmark}} \\
\cmidrule{4-6} 
 & (2) LAP Word Usage & & East Asians use significantly more low-arousal positive words than Americans. & \textbf{\textcolor{green!60!black}{\checkmark}} & \textbf{\textcolor{red}{\ding{55}}} \\
\bottomrule
\multicolumn{6}{p{1.25\textwidth}}{\footnotesize \textit{Note.} 
\textbf{Model Effect}: Main effect of model origin (testing the Instrumental Paradigm). 
\textbf{Language Effect}: Main effect of prompt language (testing the Substitutive Paradigm). 
Symbols indicate alignment with the \textbf{Human Results} column: 
\textbf{\textcolor{green!60!black}{\checkmark}} = AI pattern successfully replicates human cultural differences; 
\textbf{\textcolor{red}{\ding{55}}} = AI pattern contradicts to human differences; \textcolor{nullblue}{\textbf{$\bigcirc$}} = No significant difference observed (null effect).}
\end{tabular}%
}
\end{table*}

\subsection{Models}

To investigate cultural patterns across different modalities, we utilized a multi-tier model architecture consisting of \textit{Subject Models} (which performed the tasks) and two categories of \textit{Evaluator Models} (which scored visual outputs and analyzed textual responses).

\paragraph{Subject Models (Generative).}
For tasks requiring text generation, reasoning, or visual interpretation (Tasks 2, 3, 4, 6, 7, 8), we employed \textbf{GPT-4o} (OpenAI, USA) and \textbf{ERNIE 4.5} (Baidu, China). Both are state-of-the-art foundation models with strong bilingual capabilities. Notably, for image interpretation tasks, we utilized the vision-language capabilities of these models to interpret visual stimuli directly. For tasks requiring the synthesis of visual content from text prompts (Tasks 1 and 5), we utilized specialized text-to-image models: \textbf{DALL-E 3} (OpenAI) and \textbf{iRAG-1.0} (Baidu). iRAG-1.0 was selected as the representative Chinese image generation model to contrast with the American DALL-E 3.

\paragraph{Evaluator Models (Visual Scoring).}
To ensure objective evaluation of the generated images in Tasks 1 and 5, we employed two advanced Vision-Language Models (VLMs) as automated judges: \textbf{Gemini-2.5-Pro} (Google) and \textbf{Qwen-VL-Max} (Alibaba). Model selection was grounded in performance metrics as of November 10, 2025; according to community-based multimodal benchmarks (e.g., LMArena Vision Arena, MMMU, and Roboflow Playground), these systems were ranked as the top-performing Western and Chinese vision-language models, respectively.\footnote{Visual Benchmark references: \url{https://lmarena.ai/leaderboard/vision}, \url{https://mmmu-benchmark.github.io/}, and \url{https://playground.roboflow.com/ranking}.}

\paragraph{Evaluator Models (Text Analysis).}
For tasks yielding unstructured natural language responses (Tasks 2, 6, and 8), we required a robust method to extract specific variables (e.g., quantifying objects in a scene description, classifying emotion words) from open-ended text. To achieve this, we employed \textbf{DeepSeek-V3} (DeepSeek) and \textbf{Claude 4.5 Sonnet} (Anthropic) as automated analysts. These models were selected based on their status as the leading Chinese and Western general-purpose language models, respectively, for natural language understanding and complex reasoning as of November 5, 2025, according to major evaluation hubs (e.g., SWE-bench, Chatbot Arena, and LLM Stats).\footnote{Text Benchmark references: \url{https://www.swebench.com/}, \url{https://lmarena.ai/leaderboard}, and \url{https://llm-stats.com/}.} The use of dual evaluators from different cultural-developmental origins further controls for potential linguistic or cultural biases in the data extraction process.

\subsection{Data Collection and Analysis}

\paragraph{Experimental Design and Data Collection.}
We employed a 2 (Model Origin: Chinese vs. American) $\times$ 2 (Language: Chinese vs. English) between-subjects factorial design across all tasks. To determine the appropriate sample size, we followed the power analysis procedure outlined by \citet{lu2025cultural}, utilizing G*Power to ensure sufficient power (80\%) to detect small-to-medium effects ($d = 0.4$). Accordingly, we collected 100 independent iterations for each condition per task (Total $N = 400$ observations per task), exceeding standard benchmarks in recent AI psychology research. Data collection was automated via official APIs for all models. To capture the natural variability and generative diversity of the models, the temperature parameter was set to 1 for all tasks.

\paragraph{Validity of LLM-Based Methodology.}
The use of LLMs as both research subjects and automated evaluators is grounded in recent methodological advances in computational social science. Regarding \textit{LLM-as-Subject}, \citet{aher2023using} demonstrated that LLMs can effectively simulate human populations and replicate classic psycholinguistic findings, validating their use as proxies for studying cultural cognition. We set the temperature parameter to 1.0 to capture the probabilistic distribution of the models' responses, treating them as stochastic agents rather than deterministic tools. Regarding \textit{LLM-as-Judge}, \citet{zheng2023judging} established that strong foundation models (such as GPT-4 and Claude) achieve high agreement with human experts in open-ended evaluation tasks, often matching or exceeding inter-human reliability. Given the scale of our dataset ($N=3200$ total observations), automated evaluation provides a consistent and reproducible scoring standard. Importantly, our inferential tests are not intended to estimate human population parameters, but to characterize \textit{structured differences between conditional output distributions} under controlled prompts and sampling settings (temperature fixed at 1.0). In this sense, statistical significance is used as a descriptive tool for detecting reliable distributional shifts across experimental conditions, rather than as evidence of human-like psychological causality.

\paragraph{Statistical Analysis.} Data analysis was conducted using R (Version 4.3.1). 
For tasks yielding continuous or ordinal data (Tasks 1, 2, 3, 4, 6, 7, and 8), statistical comparisons focused on the planned contrasts defined by our experimental framework. 
Specifically, we employed independent-samples $t$-tests to examine the differences between Model Origins (Chinese vs. US) and Prompt Languages (Chinese vs. English), reporting Cohen's $d$ as the measure of effect size. 
To explicitly test for the interdependence of these factors, distinct 2 (Model) $\times$ 2 (Language) Analyses of Variance (ANOVAs) were conducted, with significant interactions decomposed using simple effects analyses (as observed in Task 3). 
Note that for Task 6 (Emotion Interpretation), separate analyses were performed for each emotion category dimension to isolate specific perception biases. 
For tasks involving categorical outcome distributions (Task 5), we employed Pearson's Chi-square tests of independence to evaluate distributional differences in generated content frequencies.

\section{Results}

The results are organized into two main domains: Cognitive Style (Tasks 1--4) and Affective Style (Tasks 5--8). For each task, we examined the main effects of Model (testing the Instrumental Paradigm) and Language (testing the Substitutive Paradigm), as well as their interaction. A summary of all tasks, including human benchmarks and the consistency of AI performance with these benchmarks, is provided in Table~\ref{tab:results_summary}.

% ==========================================================================================
% PART 1: INSTRUMENTAL PARADIGM (MODEL ORIGIN)
% ==========================================================================================
\subsection{Testing the Instrumental Paradigm: The Role of Model Origin}

\begin{figure*}[t!]
  \centering
  \includegraphics[width=\textwidth]{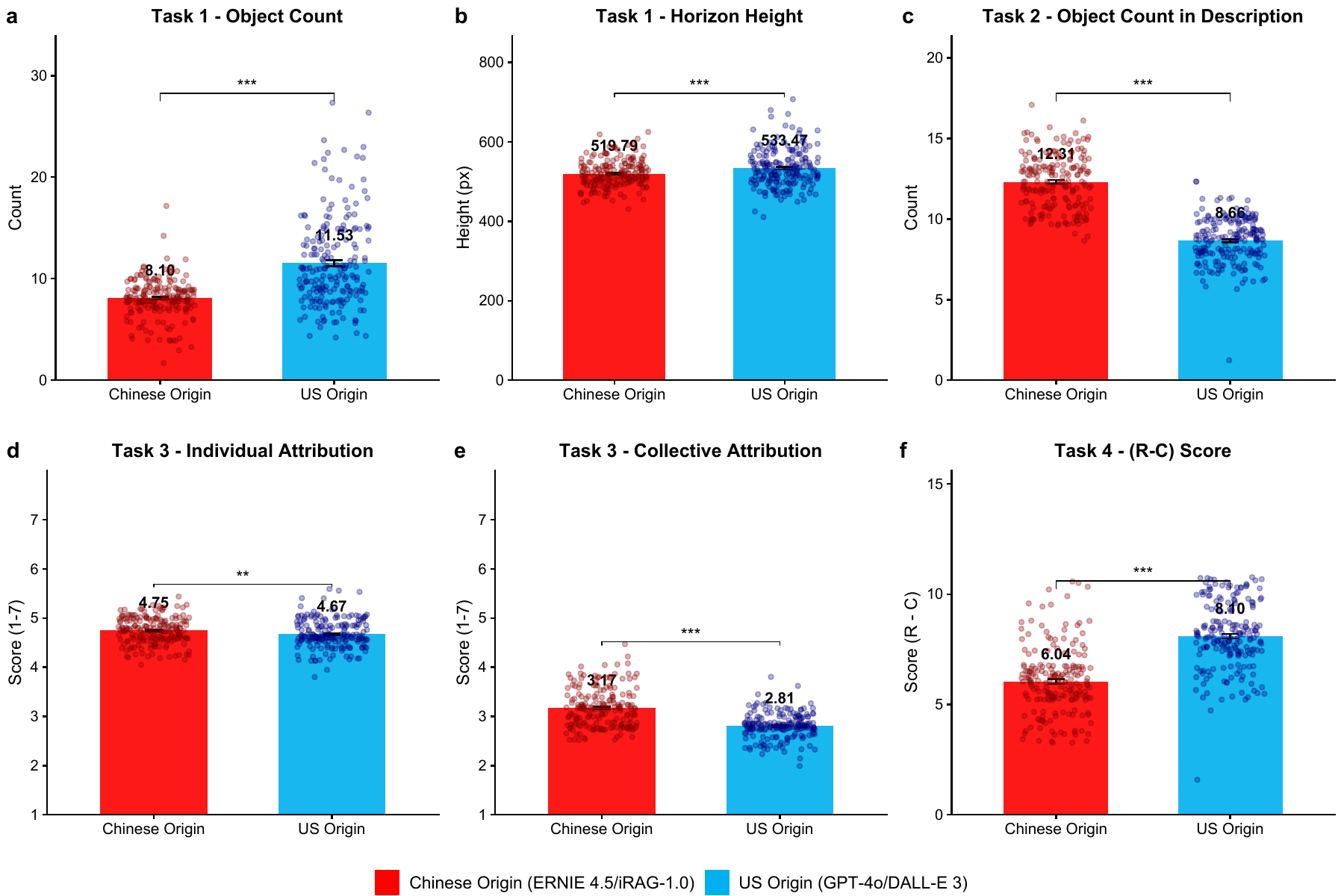}
  \caption{Results of Model Origin effects comparing Chinese-developed models (Red: ERNIE 4.5/iRAG-1.0) vs. US-developed models (Blue: GPT-4o/DALL-E 3). Data are collapsed across prompt languages to isolate the intrinsic ``cultural alignment'' of the models.
  (a) Task 1: Object count in generated landscapes.
  (b) Task 1: Horizon height in generated landscapes.
  (c) Task 2: Object count in scene descriptions.
  (d) Task 3: Individual Attribution scores.
  (e) Task 3: Collective Attribution scores.
  (f) Task 4: Relational vs. Categorical (R-C) scores.
  Note the frequent inversion of expected cultural patterns (e.g., US models scoring higher on holistic markers in Tasks 1 and 4). Significance levels: *** $p < .001$, ** $p < .01$, * $p < .05$.}
  \label{fig:cognitive_style_model}
\end{figure*}

To test the Instrumental Paradigm---the hypothesis that a model's ``cultural identity'' is determined by its developers and training corpora---we examined the main effect of Model Origin (Chinese vs. US). Results indicate that model origin predicts behavioral variances, but the directionality frequently contradicts established human cultural patterns (see Figure~\ref{fig:cognitive_style_model}).

\subsubsection{Cognitive Style}
\paragraph{Visual Attention and Construction (Tasks 1 \& 2).}
In the image generation domain (Task 1), the US model (DALL-E 3) exhibited significantly stronger ``holistic'' markers than the Chinese model (iRAG-1.0). The US model generated more objects ($M = 11.53, SD = 4.39$) than the Chinese model ($M = 8.10, SD = 1.91$), $t = -10.14, p < .001, d = 1.01, 95\% \text{ CI } [-4.09, -2.77]$ (Figure~\ref{fig:cognitive_style_model}a), and placed horizons significantly higher ($M = 533.47, SD = 46.78$) than the Chinese model ($M = 519.79, SD = 33.34$), $t = -3.37, p < .001, d = 0.34, 95\% \text{ CI } [-21.64, -5.72]$ (Figure~\ref{fig:cognitive_style_model}b). These findings are an inversion of human data, where East Asian cultural products typically feature higher complexity and higher horizons.

In contrast, for text-based description (Task 2), the Chinese model produced descriptions with significantly higher object counts ($M = 12.31, SD = 1.63$) compared to the US model ($M = 8.66, SD = 1.37$), $t = 24.22, p < .001, d = 2.42, 95\% \text{ CI } [3.36, 3.94]$ (Figure~\ref{fig:cognitive_style_model}c). While this aligns with East Asian attention to detail, the discrepancy between Task 1 (Image Generation) and Task 2 (Image Interpretation) within the same cultural models suggests a lack of cross-modal coherence.

\paragraph{Causal Attribution and Categorization (Tasks 3 \& 4).}
The Attribution task (Task 3) revealed a complex pattern. While the Chinese model scored higher on Collective Attribution ($M = 3.17, SD = 0.38$) than the US model ($M = 2.81, SD = 0.25$), consistent with collectivist norms ($t = 11.35, p < .001, d = 1.14, 95\% \text{ CI } [0.30, 0.42]$, Figure~\ref{fig:cognitive_style_model}e), it unexpectedly also scored significantly higher on Individual Attribution ($M = 4.75, SD = 0.27$) than the US model ($M = 4.67, SD = 0.28$), $t = 2.70, p = .007, d = 0.27, 95\% \text{ CI } [0.03, 0.13]$ (Figure~\ref{fig:cognitive_style_model}d). This simultaneous elevation of opposing attributional styles contradicts the trade-off typically observed in human psychology.

Finally, the Categorization task (Task 4) provided the most striking evidence against simple cultural alignment. The US model demonstrated a significantly stronger preference for relational pairing (higher R-C scores: $M = 8.10, SD = 1.44$) compared to the Chinese model ($M = 6.04, SD = 1.54$), $t = -13.81, p < .001, d = 1.38, 95\% \text{ CI } [-2.34, -1.78]$ (Figure~\ref{fig:cognitive_style_model}f). This result is diametrically opposed to the well-documented East Asian preference for relational categorization, further challenging the assumption that ``Chinese models'' inherently encode ``Chinese cognition.''

\subsubsection{Affective Style}

\begin{figure*}[t!]
  \centering
  \includegraphics[width=\textwidth]{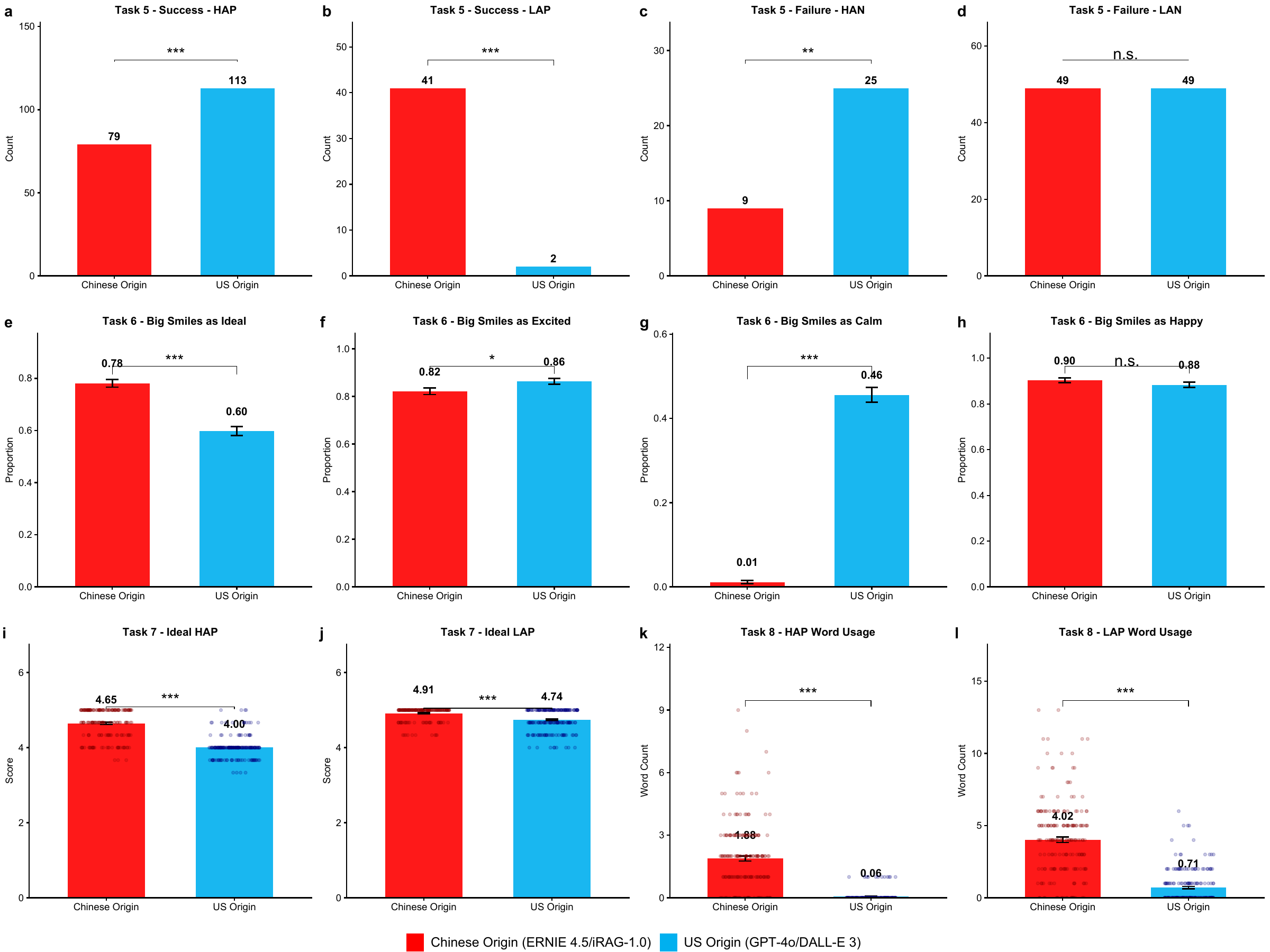}
  \caption{Results of Model Origin effects on Affective Style (Tasks 5--8). Red bars represent Chinese-developed models (ERNIE 4.5/iRAG-1.0); Blue bars represent US-developed models (GPT-4o/DALL-E 3).
  \textbf{(a--d) Task 5 (Emotion Generation):} Counts of generated emotion categories in success (a, b) and failure (c, d) scenarios.
  \textbf{(e--h) Task 6 (Emotion Interpretation):} Proportion of selecting faces as Ideal, Excited, Calm, or Happy.
  \textbf{(i--j) Task 7 (Ideal Affect):} Ratings for Ideal High Arousal Positive (HAP) and Low Arousal Positive (LAP) states.
  \textbf{(k--l) Task 8 (Ideal State):} Frequency of HAP and LAP word usage in open-ended descriptions.
  Significance levels: *** $p < .001$, ** $p < .01$, * $p < .05$, n.s. not significant.}
  \label{fig:affective_style_model}
\end{figure*}

In the domain of affect, we observed a striking dissociation between generative imagery and textual reasoning. While image generation adhered to cultural stereotypes, text-based tasks revealed a ``hyper-emotionality'' in the Chinese model that contradicted established human patterns (see Figure~\ref{fig:affective_style_model}).

\paragraph{Generative Emotional Expression (Task 5).}
In the generative domain, model behavior largely aligned with the Instrumental Paradigm and human cultural norms. In success scenarios, the US model generated significantly more High Arousal Positive (HAP) content ($N=113$) compared to the Chinese model ($N=79$), $\chi^2 = 28.36, p < .001$. Conversely, the Chinese model generated significantly more Low Arousal Positive (LAP) content ($N=41$) than the US model ($N=2$), $\chi^2 = 40.91, p < .001$ (Figure~\ref{fig:affective_style_model}a--b). Similarly, in failure scenarios, the US model exhibited a significantly stronger preference for High Arousal Negative (HAN) content ($N=25$) compared to the Chinese model ($N=9$), $\chi^2 = 7.71, p = .005$, mirroring the Western tendency to externalize distress (Figure~\ref{fig:affective_style_model}c--d).

\paragraph{Emotion Interpretation and Ideal Affect (Tasks 6--8).}
However, interpretive and text-based tasks revealed results that diametrically opposed the cultural hypothesis. In the Emotion Interpretation task (Task 6), the Chinese model was significantly \textit{more} likely to perceive ``big smiles'' as Ideal ($M=0.78$) compared to the US model ($M=0.60$), $t(1553.54) = 8.10, p < .001, d = 0.40$ (Figure~\ref{fig:affective_style_model}e), an inversion of the American cultural preference for high-intensity smiles.

This pattern of inversion intensified in text-based preferences. For Ideal Affect (Task 7), the Chinese model rated Ideal HAP significantly higher ($M=4.65$) than the US model ($M=4.00$), $t(368.29) = 18.52, p < .001, d = 1.85$ (Figure~\ref{fig:affective_style_model}i). Even more strikingly, in the open-ended Ideal State task (Task 8), the Chinese model utilized significantly more HAP-related words ($M=1.88$) than the US model ($M=0.06$), $t(207.26) = 15.42, p < .001, d = 1.54$ (Figure~\ref{fig:affective_style_model}k).

Crucially, this was not a trade-off; the Chinese model also scored significantly higher on LAP metrics in both Task 7 ($t = 7.09, p < .001$) and Task 8 ($t = 16.12, p < .001$). These findings suggest that the Chinese model (ERNIE 4.5) exhibits a generalized ``affective intensity'' or verbosity in text processing, elevating both high- and low-arousal values simultaneously, rather than encoding a specific, culturally coherent affective profile.

\subsection{Testing the Substitutive Paradigm: The Role of Prompt Language}

\begin{figure*}[t!]
  \centering
  \includegraphics[width=\textwidth]{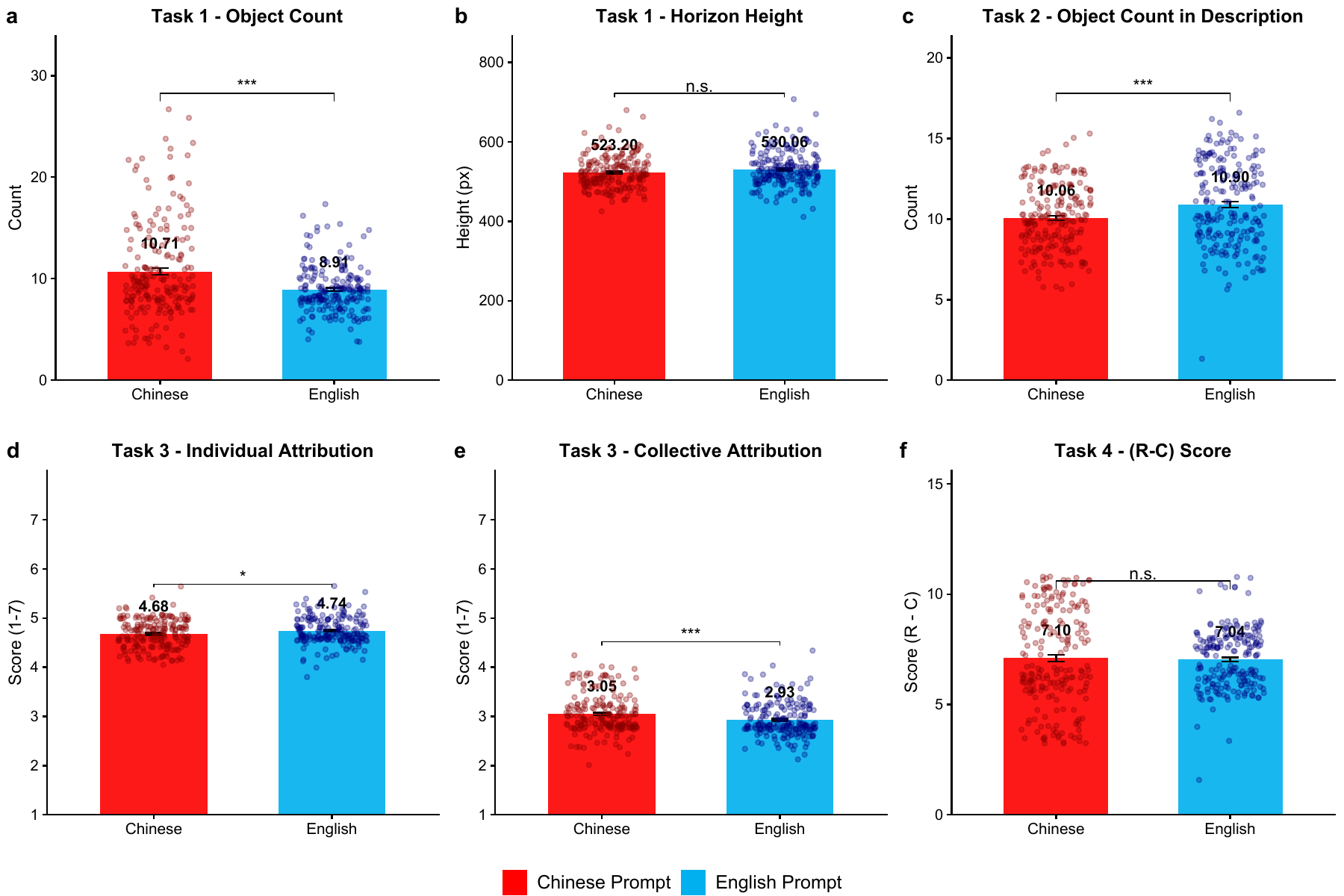}
  \caption{Results of Cognitive Style tasks (Tasks 1--4) comparing the effects of Chinese (Red) vs. English (Blue) prompt languages. 
  (a) Task 1: Object count in generated landscapes, serving as a proxy for holistic attention.
  (b) Task 1: Horizon height in generated landscapes.
  (c) Task 2: Total object count in scene descriptions.
  (d) Task 3: Individual Attribution scores (1--7 scale).
  (e) Task 3: Collective Attribution scores (1--7 scale).
  (f) Task 4: Relational vs. Categorical (R-C) categorization scores.
  Bars represent group means; error bars indicate standard error of the mean ($\pm$SEM). Individual data points are overlaid to show distribution. Significance levels: *** $p < .001$, ** $p < .01$, * $p < .05$, n.s. not significant ($p > .05$).}
  \label{fig:cognitive_style_language}
\end{figure*}

Having established that model origin does not deterministically encode cultural patterns as the Instrumental Paradigm suggests, we next examine whether cultural patterns are instead dynamically activated by the language of interaction, as proposed by the Substitutive Paradigm. To evaluate this hypothesis---that prompt language triggers culture-specific cognitive modes---we analyzed the main effect of Language (Chinese vs. English). Data were collapsed across model origins to isolate the specific influence of linguistic priming. Visual results are presented in Figure~\ref{fig:cognitive_style_language}.

\subsubsection{Cognitive Style}

\paragraph{Visual Attention and Construction (Tasks 1 \& 2).}
Results from the image-based tasks revealed a significant dissociation between content generation and spatial composition. In the Landscape Generation task (Task 1), prompt language successfully modulated the quantity of contextual elements. Consistent with East Asian holistic cognition, models prompted in Chinese generated significantly more objects ($M = 10.71, SD = 4.70$) than those prompted in English ($M = 8.91, SD = 2.26$), $t(286.19) = 4.88, p < .001, d = 0.49, 95\% \text{ CI } [1.07, 2.53]$ (see Figure~\ref{fig:cognitive_style_language}a). However, this cultural alignment did not extend to spatial framing. There was no significant difference in horizon height between Chinese ($M = 523.20, SD = 39.72$) and English ($M = 530.06, SD = 42.35$) conditions, $t(396.37) = -1.67, p = .096, d = 0.17, 95\% \text{ CI } [-14.93, 1.21]$ (Figure~\ref{fig:cognitive_style_language}b), failing to replicate the human tendency for East Asians to utilize higher horizons.

Furthermore, the Scene Interpretation task (Task 2) yielded results directly contradictory to the cultural hypothesis. While human data suggests East Asians (and by extension, Chinese prompts) should elicit more detailed background descriptions, we observed the opposite: English prompts elicited significantly higher object counts in descriptions ($M = 10.90, SD = 2.61$) than Chinese prompts ($M = 10.06, SD = 2.02$), $t(374.94) = -3.60, p < .001, d = 0.36, 95\% \text{ CI } [-1.30, -0.38]$ (Figure~\ref{fig:cognitive_style_language}c). This suggests that verbose description in LLMs may be a feature of English training data density rather than an activation of holistic attention.

\paragraph{Causal Attribution and Categorization (Tasks 3 \& 4).}
In the text-based reasoning domains, the effectiveness of linguistic priming was similarly inconsistent. The Attribution task (Task 3) provided the strongest support for the Substitutive Paradigm. When prompted in Chinese, models scored significantly lower on Individual Attribution ($M = 4.68, SD = 0.28$) compared to English prompts ($M = 4.74, SD = 0.27$), $t(397.51) = -2.33, p = .021, d = 0.23, 95\% \text{ CI } [-0.12, -0.01]$ (Figure~\ref{fig:cognitive_style_language}d). Conversely, Chinese prompts elicited significantly higher Collective Attribution scores ($M = 3.05, SD = 0.39$) than English prompts ($M = 2.93, SD = 0.34$), $t(390.68) = 3.37, p < .001, d = 0.34, 95\% \text{ CI } [0.05, 0.19]$ (Figure~\ref{fig:cognitive_style_language}e). Although the effect sizes were small, the directionality perfectly mirrors human cross-cultural differences.

However, this pattern collapsed in the Categorization task (Task 4). Despite the robust human finding that East Asian languages promote relational pairing over categorical pairing, we found no significant difference in Relational-Categorical (R-C) scores between Chinese ($M = 7.10, SD = 2.15$) and English ($M = 7.04, SD = 1.40$) conditions, $t(342.66) = 0.33, p = .741, d = 0.03, 95\% \text{ CI } [-0.30, 0.42]$ (Figure~\ref{fig:cognitive_style_language}f). The prompt language failed to prime the expected ``system of thought,'' suggesting that the conceptual relationships between these terms are likely stabilized in the models' multilingual embedding space regardless of the input language.

\subsubsection{Affective Style}

\begin{figure*}[t!]
  \centering
  \includegraphics[width=\textwidth]{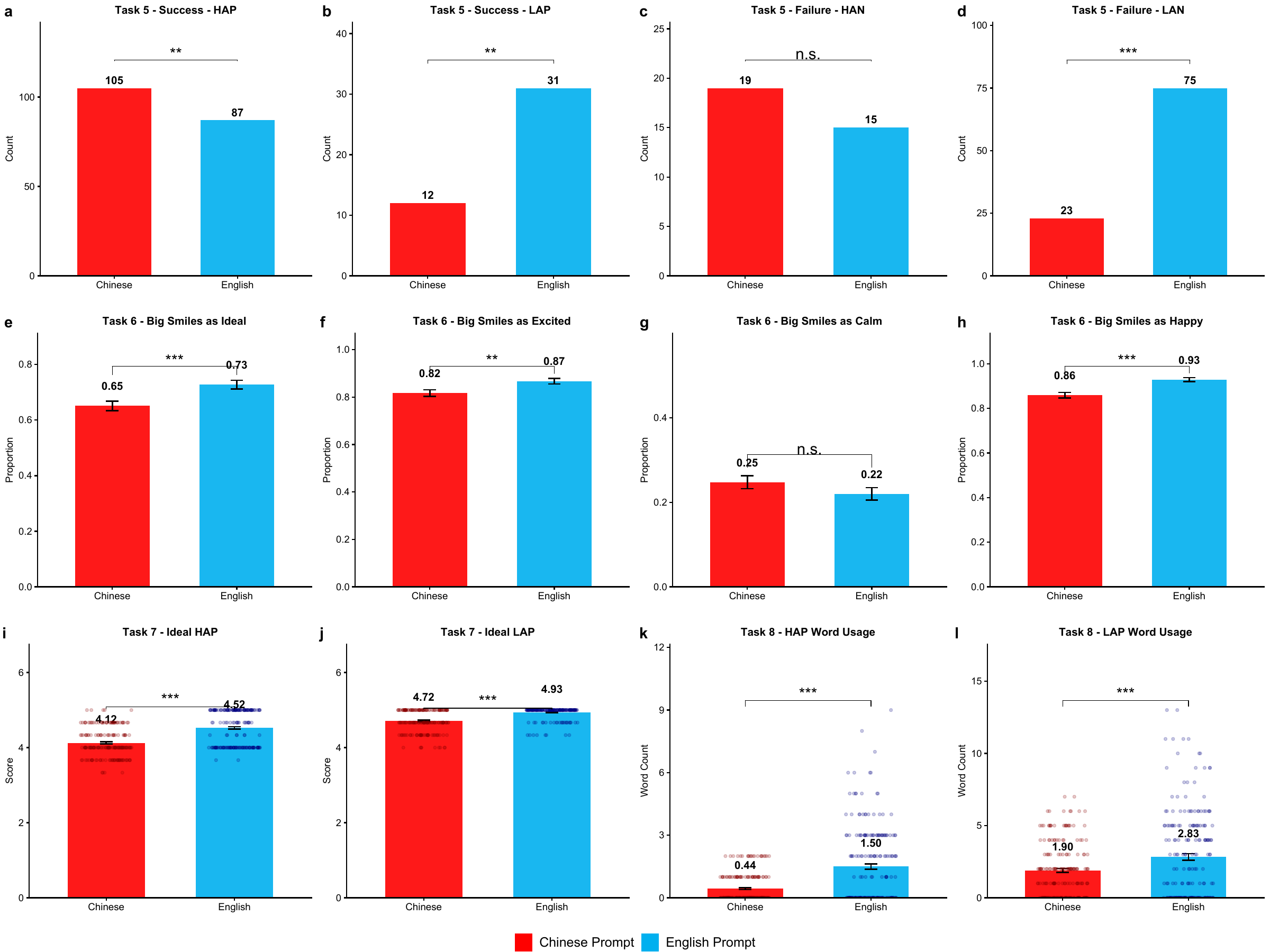}
  \caption{Results of Prompt Language effects on Affective Style (Tasks 5--8). Red bars represent Chinese prompts; Blue bars represent English prompts.
  \textbf{(a--d) Task 5:} Effects of language on generated emotion categories (HAP, LAP, HAN, LAN).
  \textbf{(e--h) Task 6:} Effects of language on facial expression interpretation.
  \textbf{(i--j) Task 7:} Impact of language on Ideal HAP and LAP ratings.
  \textbf{(k--l) Task 8:} Impact of language on emotional word count in ideal state descriptions.
  Note the ``Intensity Bias'' in text-based tasks (i--l), where English prompts frequently elicit higher values across both high-arousal and low-arousal metrics simultaneously compared to Chinese prompts.
  Significance levels: *** $p < .001$, ** $p < .01$, * $p < .05$, n.s. not significant.}
  \label{fig:affective_style_language}
\end{figure*}

\paragraph{Generative Emotional Expression (Task 5).}
In the generative domain, prompt language significantly influenced the emotional content of generated images, though the directionality was complex. Chi-square tests revealed significant linguistic dependencies for both High Arousal Positive (HAP) content ($\chi^2 = 7.53, p = .006$) and Low Arousal Positive (LAP) content ($\chi^2 = 9.18, p = .002$). Similarly, in failure scenarios, Low Arousal Negative (LAN) content differed significantly by language ($\chi^2 = 44.86, p < .001$), while High Arousal Negative (HAN) content did not ($\chi^2 = 0.31, p = .579$; see Figure~\ref{fig:affective_style_language}a--d).

\paragraph{Emotion Interpretation and Ideal Affect (Tasks 6--8).}
When interpreting facial expressions, English prompts consistently biased models toward high-arousal interpretations (see Figure~\ref{fig:affective_style_language}). Models prompted in English were significantly more likely to identify ``big smiles'' as \textit{Ideal} ($M = 0.73$) compared to Chinese prompts ($M = 0.65$), $t(1590.67) = -3.31, p < .001, d = 0.17$. Similarly, English prompts elicited higher rates of perceiving big smiles as \textit{Excited} ($M = 0.87$ vs. $M = 0.82$, $t(1571.58) = -2.75, p = .006, d = 0.14$) and \textit{Happy} ($M = 0.93$ vs. $M = 0.86$, $t(1470.83) = -4.57, p < .001, d = 0.23$). No significant difference was found for the \textit{Calm} dimension ($t = 1.30, p = .194$).

The text-based affective tasks provided the clearest evidence of an ``intensity bias'' inherent to the English language in LLMs. For Ideal Affect (Task 7), English prompts elicited significantly higher ratings for Ideal HAP ($M = 4.52$) than Chinese prompts ($M = 4.12$), $t(381.66) = -9.30, p < .001, d = 0.93$. However, contradicting the cross-cultural expectation that East Asian languages should promote higher Ideal LAP, English prompts \textit{also} produced significantly higher Ideal LAP ratings ($M = 4.93$) than Chinese prompts ($M = 4.72$), $t(310.00) = -9.17, p < .001, d = 0.92$.

This pattern of simultaneous elevation was robustly replicated in the open-ended Ideal State task (Task 8). Models prompting in English generated significantly more HAP-related words ($M = 1.50$) than those in Chinese ($M = 0.44$), $t(251.33) = -7.61, p < .001, d = 0.76$. Yet, they also generated significantly more LAP-related words ($M = 2.83$ vs. $M = 1.90$), $t(323.06) = -3.61, p < .001, d = 0.36$. 

These results demonstrate that English prompts do not simply activate a ``Western cultural mind'' (which would favor HAP over LAP), but rather trigger a generalized mode of emotional verbosity that elevates affective expression across the arousal spectrum.

% ==========================================================================================
% PART 3: EMERGENT MECHANISMS (NEW SECTION AT THE END OF RESULTS)
% ==========================================================================================

\subsection{Emergent Mechanisms: Evidence for Machine Culture}

The analyses above demonstrate that neither Model Origin nor Prompt Language provides a consistent account of LLM behavior, with results often contradicting human cultural patterns. To investigate the underlying mechanisms driving these inconsistencies, we conducted a targeted post-hoc analysis. This revealed two distinct phenomena that distinguish LLM behavior from human cognition: \textit{Reversal/Asymmetry} and \textit{Service Persona Camouflage}.

\subsubsection{Phenomenon I: Reversal and Asymmetry in Language Priming}

\begin{figure*}[t!]
  \centering
  \includegraphics[width=\textwidth]{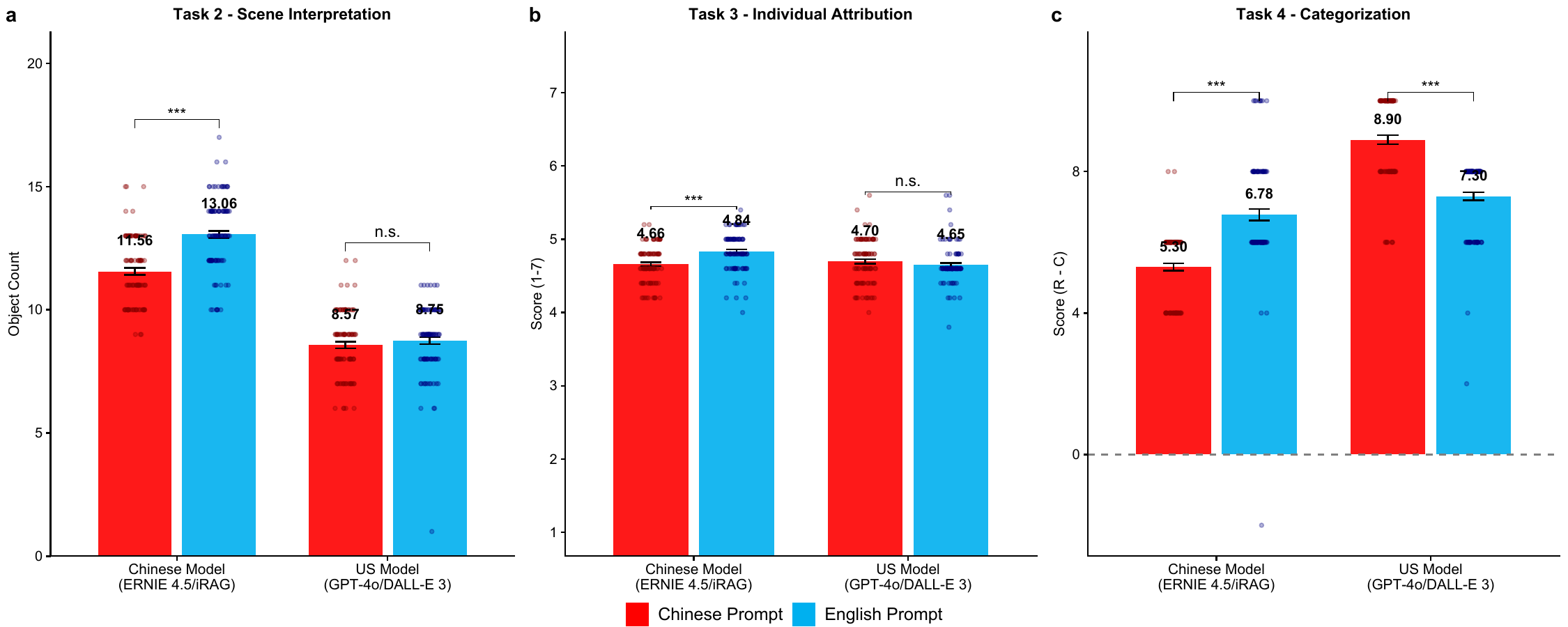}
  \caption{Evidence of \textbf{Reversal and Asymmetry in Language Priming} across cognitive tasks. 
  (a) \textbf{Task 2 (Scene Interpretation)}: Contrary to human cultural patterns, English prompts elicited significantly higher object counts (more holistic) than Chinese prompts across models.
  (b) \textbf{Task 3 (Individual Attribution)}: An interaction effect reveals functional asymmetry; Chinese prompts significantly reduced individual attribution for the Chinese model but had no significant effect on the US model.
  (c) \textbf{Task 4 (Categorization)}: A striking reversal where English prompts elicited significantly higher relational scoring (R-C) in the Chinese model, contradicting the expected East Asian preference for relational thinking in native language contexts.
  Significance levels: *** $p < .001$, ** $p < .01$, * $p < .05$, n.s. not significant.}
  \label{fig:reversal_asymmetry}
\end{figure*}

While the Substitutive Paradigm posits that language activates cultural schemas, our granular analysis reveals that language priming in LLMs often triggers \textit{reverse} or \textit{asymmetric} effects (Figure~\ref{fig:reversal_asymmetry}).

First, we observed a \textbf{Cultural Reversal} where English prompts triggered ``East Asian'' cognitive traits more strongly than Chinese prompts. In Task 2 (Scene Interpretation), English prompts elicited significantly higher object counts ($M=10.91$) than Chinese prompts ($M=10.07$, $t=-3.60, p<.001$), suggesting that verbosity in English training data may override cultural holistic tendencies (Figure~\ref{fig:reversal_asymmetry}a). Similarly, in Task 4 (Categorization), the Chinese Model exhibited a reversal: it scored significantly \textit{lower} on relational thinking when prompted in Chinese ($M=5.30$) compared to English ($M=6.78$, $t=-7.74, p<.001$), directly contradicting human bilingual evidence (Figure~\ref{fig:reversal_asymmetry}c).

Second, we identified \textbf{Functional Asymmetry}. In Task 3 (Attribution), a significant Model $\times$ Language interaction ($F(1, 396) = 18.21, p < .001$) indicated that while the Chinese model was sensitive to linguistic priming (shifting towards individual attribution in English, $t=-4.96$), the US model remained unresponsive ($t=1.26, n.s.$). This suggests that language does not function as a universal cultural key, but interacts idiosyncratically with model architecture (Figure~\ref{fig:reversal_asymmetry}b).

\subsubsection{Phenomenon II: Service Persona Camouflage}

\begin{figure*}[t!]
  \centering
  \includegraphics[width=\textwidth]{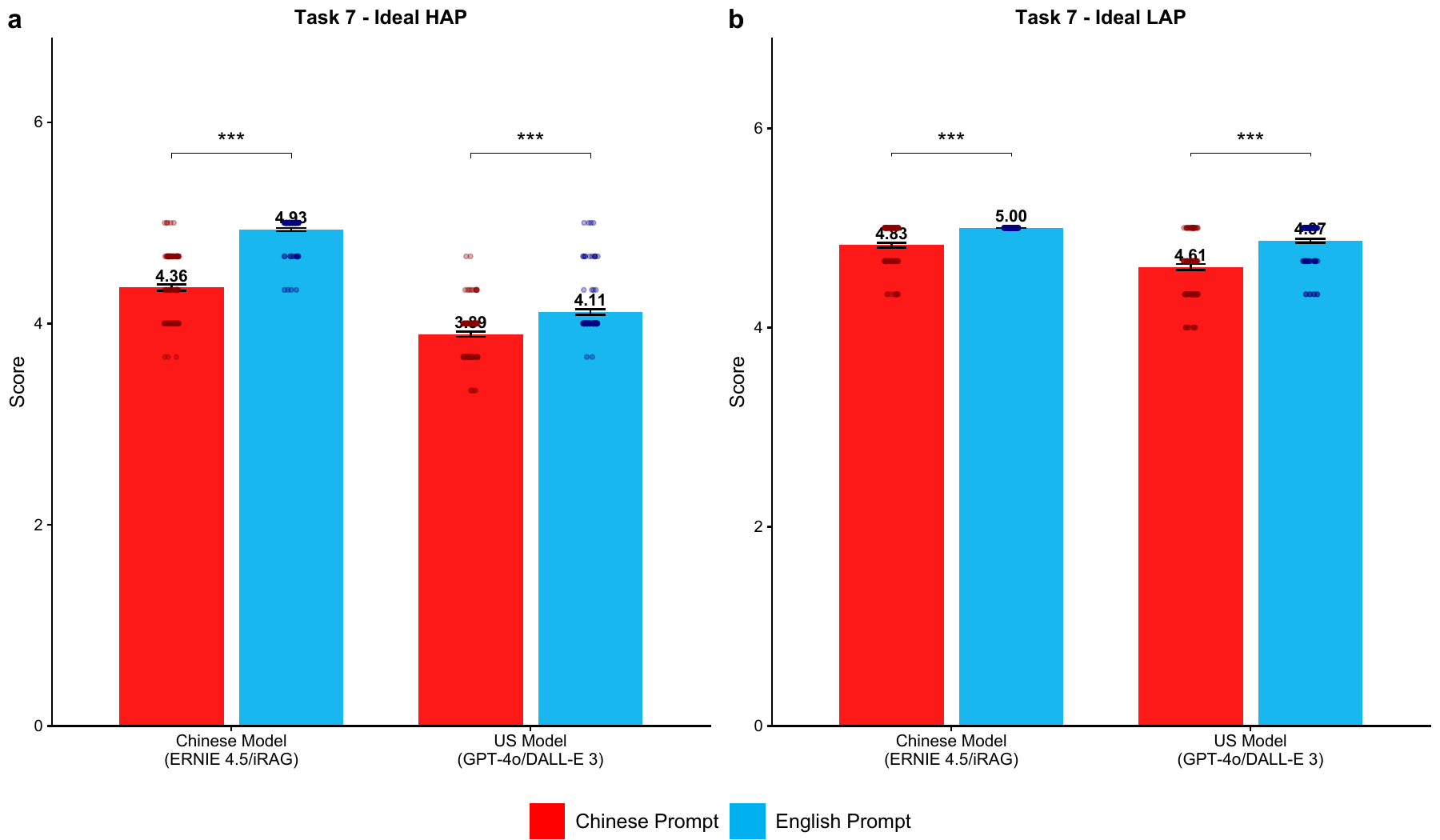}
  \caption{Evidence of \textbf{Service Persona Camouflage} in Ideal Affect (Task 7). 
  (a) Ideal High Arousal Positive (HAP) scores. 
  (b) Ideal Low Arousal Positive (LAP) scores. 
  Note the extreme ceiling effects and lack of variance, particularly in the Chinese Model conditioned with English Prompts (Panel b, right bar), which exhibited a perfect mean of 5.00 with $SD=0.00$. This indicates a collapse of cultural variation into a hyper-positive, alignment-driven ``service persona'' rather than a simulation of human affective diversity.
  Significance markers indicate interaction effects or simple main effects.}
  \label{fig:camouflage}
\end{figure*}

Further challenging the anthropomorphic view is the phenomenon of \textit{Service Persona Camouflage}, observed most clearly in the Ideal Affect task (Task 7). Unlike humans, who trade off between High and Low Arousal positive affect, LLMs exhibited a generalized ``positivity bias'' with extreme ceiling effects (Figure~\ref{fig:camouflage}).

Statistical analysis revealed a collapse of variance inconsistent with human psychological data. The most extreme manifestation occurred in the Chinese Model conditioned with English prompts: for Ideal Low Arousal Positive (LAP) affect, the model produced a perfect mean score of $M=5.00$ with a standard deviation of $SD=0.00$ and a $100\%$ ceiling rate. This zero-variance result suggests the model is not simulating a cultural preference for calmness, but rather enacting a rigid, alignment-driven ``service persona'' that seeks to maximize all positive attributes simultaneously (Figure~\ref{fig:camouflage}b). This ``camouflage'' effectively masks any latent cultural tendencies, rendering the model's outputs a reflection of RLHF (Reinforcement Learning from Human Feedback) tuning rather than cultural cognition.

\section{Discussion}

The present study offers a systematic evaluation of cultural patterns in Large Language Models, revealing a critical dissociation between human cultural frameworks and model behavior. Our findings challenge the prevailing \textit{Instrumental} and \textit{Substitutive} paradigms, suggesting that LLMs do not merely reflect the culture of their developers nor simulate the cultural cognition of human users. Instead, they exhibit an emergent \textbf{Machine Culture} \citep{brinkmann2023machine}, characterized by functional asymmetry, cross-modal incoherence, and alignment-driven homogenization.

\subsection{The Failure of the Instrumental Paradigm: From National Repositories to Mode Collapse}
Contrary to the Instrumental Paradigm, model origin was a poor predictor of cultural alignment. While it is widely reported that the Chinese model (ERNIE) integrates massive Chinese-language datasets \citep{sun2021ernie}, our results show it frequently exhibits ``Western'' or ``Global'' traits (e.g., individualistic attribution). This apparent paradox is not incompatible with heavy Chinese-language training: in contemporary foundation models, cultural signals in language-specific corpora can be diluted by (i) substantial overlap with globally crawled web and technical corpora, and (ii) downstream alignment objectives optimized against universalized helpfulness/safety criteria rather than culturally localized norms \citep{gao2020pile, ouyang2022training, bai2022constitutional, rafailov2023direct}.

This pattern likely stems from two additional mechanisms. First, the \textit{curse of recursion} \citep{shumailov2023curse} suggests that training on large-scale web data (often partially model-generated) induces distributional \textit{mode collapse}, where distinct cultural distributions converge toward a generic mean. Second, Reinforcement Learning from Human Feedback (RLHF) creates an ``alignment tax'' \citep{casper2023open}, which can mask implicit cultural differences embedded during pre-training \citep{arora2023probing} behind a homogenized explicit output layer. Thus, what appears as a ``modernizing bias'' is better interpreted as \textit{structural convergence} under globalized data pipelines and alignment constraints, rather than an empirically demonstrated temporal evolution.

\subsection{The Failure of the Substitutive Paradigm: Prompt Sensitivity vs. Cultural Frame-Switching}
The Substitutive Paradigm posits that language activates specific cultural mindsets. Our observation of \textit{Cultural Reversal} dismantles this assumption. Unlike human bilinguals who access internalized cultural schemas \citep{hong2000multicultural}, LLMs' behavior is driven by probabilistic associations in high-dimensional space. English prompts often function as a prime for \textit{capability} and \textit{verbosity} rather than Western culture, activating deeper representations that result in more detailed (statistically ``holistic'') outputs. This instability aligns with recent findings that LLMs exhibit extreme sensitivity to prompt formatting \citep{khan2025randomness, sclar2023quantifying}, where superficial structural changes override any latent cultural simulation.

\subsection{Machine Culture: Superposition and Service Personas}
Importantly, \textbf{Machine Culture should not be understood as a psychological construct}. Rather, it is a \textit{computationally emergent phenomenon} arising from optimization objectives, representational superposition, and alignment constraints in large-scale models. In this sense, Machine Culture does not describe internalized beliefs, values, or stable mental schemas, but characterizes systematic regularities in model behavior that emerge from training dynamics and deployment conditions \citep{bommasani2021opportunities, shanahan2024talking}.

This framing directly departs from anthropomorphic interpretations of language models as possessing human-like cultural cognition. As argued by \citet{shanahan2024talking}, treating LLMs as psychological agents risks category errors that obscure their true computational nature. Instead, consistent with the foundation model perspective \citep{bommasani2021opportunities}, cultural patterns observed in LLM outputs should be understood as emergent system-level properties shaped by data scale, architectural constraints, and alignment pipelines, rather than reflections of coherent cultural minds.

This perspective also aligns with prior critiques of large language models as stochastic systems optimized over heterogeneous corpora rather than grounded communicative agents \citep{bender2021dangers}. From this viewpoint, what we term \textit{Machine Culture} captures how globally aggregated data and alignment objectives can yield culturally legible yet internally unstable behaviors, producing apparent cultural signals without corresponding human-like cultural understanding.

\section{Limitations and Future Directions}

While this study proposes a novel framework for Machine Culture, several limitations warrant future investigation. 
First, regarding \textit{model evolution}, we did not compare different versions of the same model family (e.g., ERNIE 3.0 vs. 4.0). Consequently, our discussion on the ``convergence'' of biases is a theoretical inference based on cross-sectional data rather than longitudinal evidence. 
Second, our measures focused on \textit{explicit outputs}. Recent research suggests that while RLHF may strongly align explicit responses to a globalized normative standard, implicit structures embedded during pre-training may persist beneath this aligned surface \citep{arora2023probing, casper2023open}. Future studies should therefore move beyond direct self-report–style prompts and adopt \textit{implicit testing paradigms} adapted for LLMs. 

Concretely, this could include (i) association-based probing tasks analogous to the Implicit Association Test (IAT), where response latencies, choice probabilities, or logit differences are examined under controlled contrasts; (ii) causal or representational probing of latent value dimensions using intervention or activation-based methods; and (iii) systematic dissociations between aligned explicit outputs and latent preference representations. Such approaches would allow researchers to disentangle alignment-induced surface behaviors from deeper representational biases in large language models.

Third, our results may be sensitive to \textit{prompt engineering}. We utilized single, standardized prompts for each task to maintain experimental control. However, given that LLMs demonstrate extreme sensitivity to prompt formatting \citep{khan2025randomness}, future work should employ prompt variations to test the robustness of these cultural patterns.
Finally, while we leveraged state-of-the-art LLMs as evaluators, we did not calculate human-AI inter-coder reliability for this specific dataset. Although \citet{zheng2023judging} support the validity of this approach, human verification on a subset of data would further strengthen the results.

\section{Conclusion}

This study marks a departure from anthropomorphic evaluations of AI. By identifying phenomena such as \textit{Cultural Reversal} and \textit{Service Persona Camouflage}, we demonstrate that Large Language Models represent a distinct ontology of intelligence. They are not imperfect mirrors of human culture, but sophisticated statistical artifacts that generate a unique \textbf{Machine Culture}. As these models increasingly mediate human knowledge, understanding their emergent properties—specifically how they compress, superpose, and homogenize cultural information—is no longer just a technical challenge, but a critical imperative for Cognitive Science.

\printbibliography

% ==========================================
% APPENDIX START
% ==========================================
\clearpage
\onecolumn 
\appendix

\section{Appendix: Computational Implementation Details}

This appendix details the computational pipelines used for stimulus generation and automated evaluation. The workflow is divided into three components: (A) Subject Model Generation (Tasks 1--8), (B) Visual Evaluator Pipeline (Tasks 1 \& 5), and (C) Text Evaluator Pipeline (Tasks 2, 6, \& 8).

\subsection{A. Subject Model Generation Pipeline}
To examine the effect of prompt language (Substitutive Paradigm), all tasks were administered in both English and Chinese. Table \ref{tab:bilingual_prompts} presents the core prompts used for the key tasks in both languages.

\begin{table*}[h]
\centering
\caption{Bilingual Prompts for Subject Model Generation (Excerpts)}
\label{tab:bilingual_prompts}
\small
\renewcommand{\arraystretch}{1.5}
\begin{tabular}{p{0.15\textwidth} p{0.40\textwidth} p{0.40\textwidth}}
\toprule
\textbf{Task} & \textbf{English Prompt} & \textbf{Chinese Prompt} \\
\midrule
\textbf{Task 1} \newline (Landscape Gen.) & ``Draw a landscape picture that includes at least a house, a tree, a river, a person, and a horizon. Feel free to draw additional objects...'' & ``画一幅风景画,至少包括一座房子、一棵树、一条河、一个人和一个地平线。可以随意绘制其他对象...'' \\
\midrule
\textbf{Task 3} \newline (Attribution) & ``In a particular company, a group of coworkers was responsible for completing a very important project... One coworker, `Z', consistently showed up late...'' & ``在一家单位里,有一个团队负责完成一个非常重要的项目... 组里的成员小李经常开会迟到...'' \\
\midrule
\textbf{Task 5} \newline (Emotion Gen.) & ``Imagine you are a human being. Now draw a picture illustrating a situation in the past 5 years in which you succeeded in achieving an important goal.'' & ``假设你就是一个人类。现在请画一幅图，描绘出过去五年中你成功达成一个重要目标的情景。'' \\
\midrule
\textbf{Task 7} \newline (Ideal Affect) & ``Imagine you are a person... rate, over the course of a typical week, how often in an ideal world you would want to have this feeling...'' & ``想象你是一个人... 评估在典型的一周的时间里,理想情况下,你想有这个感觉的频率是多少...'' \\
\bottomrule
\end{tabular}
\end{table*}

\subsection{B. Visual Evaluator Pipeline (Scoring Models)}
Visual outputs were scored using \textbf{Gemini-2.5-Pro} and \textbf{Qwen-VL-Max}. The evaluator prompts were also matched to the language of the stimulus to ensure cultural consistency during evaluation.

\subsubsection{B.1. Cognitive Style Evaluation (Task 1)}
We used the following system prompts to quantify holistic cognitive markers.

\begin{lstlisting}[language=Python, caption={System Prompt for Visual Scoring (Chinese Version)}]
IMAGE_EVALUATION_PROMPT_CHINESE = """
请分析这张图片并完成以下两项任务：

1. 物体计数：数出图片中除了【一座房子、一棵树、一条河、一个人】之外的额外物体数量。
   - 只计算明确可辨认的物体
   - 不要重复计数同一物体
   
2. 地平线高度：计算地平线距离图片下端的像素距离。
   - 图片尺寸为1024x1024像素
   - 以图片下端为0起始线
   - 结果为整数（单位：像素）

请只返回JSON格式的结果：{"object_count": 数字, "horizon_height": 整数}
"""
\end{lstlisting}

\subsubsection{B.2. Affective Style Evaluation (Task 5)}
The following prompt was used to classify generated emotions into Affect Valuation Theory categories (HAP/LAP).

\begin{lstlisting}[language=Python, caption={System Prompt for Visual Affective Scoring (English Version)}]
IMAGE_EVALUATION_PROMPT_ENGLISH = """
Please encode the valence and arousal of this image.

Valence evaluation rules:
- Images primarily expressing pleasant emotions: "positive"
- Images primarily expressing unpleasant emotions: "negative"

Arousal evaluation rules:
- Images showing high energy levels: "high arousal"
- Images showing low energy levels: "low arousal"

Please only return the result in JSON format: 
{"valence": "positive/negative/neutral", "arousal": "high arousal/low arousal"}
"""
\end{lstlisting}

\subsection{C. Text Evaluator Pipeline (Analysis Models)}
For open-ended text tasks (Tasks 2, 6, 8), \textbf{DeepSeek-V3} and \textbf{Claude 4.5 Sonnet} were employed as automated raters.

\subsubsection{C.1. Holistic Interpretation Scoring (Task 2)}
This prompt quantified ``verbosity'' and ``holistic attention'' in textual descriptions.

\begin{lstlisting}[language=Python, caption={Evaluator Prompt for Task 2 (Chinese Version)}]
SYSTEM_PROMPT_CHINESE = """
你是一个专业的文本分析助手。你的任务是分析给定的图片描述文本。

任务说明：
文本中描述了一个场景，其中"最前面的三条大鱼"是主要对象。
你需要做的：
统计文本中除了"最前面的三条大鱼"之外，还提到了多少个背景物体（包括动物、植物、物品等）。

重要：
1. 只计数背景物体的数量
2. 以JSON格式输出：{"background_object_count": 数字}
"""
\end{lstlisting}

\subsubsection{C.2. Affective Word Statistics (Task 8)}
This prompt extracted the frequency of emotional words from ``Ideal State'' descriptions.

\begin{lstlisting}[language=Python, caption={Evaluator Prompt for Task 8 (English Version)}]
SYSTEM_PROMPT_ENGLISH = """
You are a professional text emotion analysis assistant. 
The given content is a descriptive text about an "ideal state".

What you need to count:
1. Total number of emotional words.
2. Number of high-arousal positive emotion words (e.g., excited, thrilled).
3. Number of low-arousal positive emotion words (e.g., calm, peaceful).

Output in JSON format: 
{"total_emotion_words": number, "high_arousal_positive": number, "low_arousal_positive": number}
"""
\end{lstlisting}

\end{document}